 \documentclass[final,1p,times]{elsarticle}

\usepackage{graphicx}
\usepackage{amssymb}
\usepackage{amsmath}
\usepackage{algpseudocode}
\usepackage{algorithm}
\usepackage{todonotes}
\usepackage{rotating}
\usepackage{multirow}
\usepackage{array}
\usepackage{tabto}
\usepackage{color, soul}
\usepackage{url}
\usepackage[colorlinks]{hyperref}
\usepackage{cleveref}

\begin{document}

\begin{frontmatter}





 

\title{A curated collection of COVID-19 online datasets}
%
%
\author{Isa Inuwa-Dutse $\hspace{20mm}$  Ioannis Korkontzelos \\
\textit{I.Inuwa-Dutse@herts.ac.uk $\hspace{2mm}$ Yannis.Korkontzelos@edgehill.ac.uk}}

%
\begin{abstract}
One of the defining moments of the year 2020 is the outbreak of Coronavirus Disease (Covid-19), a deadly virus affecting the body's respiratory system to the point of needing a breathing aid via ventilators. As of June 21, 2020 there are 12,929,306 confirmed cases and 569,738 confirmed deaths across 216 countries, areas or territories. 
The scale of spread and impact of the pandemic left many nations grappling with preventive and curative approaches. The infamous \textit{lockdown measure} introduced to mitigate the virus spread has altered many aspects of our social routines in which demand for online-based services skyrocketed. As the virus propagate, so does misinformation and fake news around it via online social media, which seems to favour virality over veracity. With a majority of the populace confined to their homes for a long period, vulnerability to the toxic impact of online misinformation is high. A case in point is the various myths and disinformation associated with the Covid-19, which, if left unchecked, could lead to a catastrophic outcome and hamper the fight against the virus. 
 
While the scientific community is actively engaged in identifying the virus treatment, there is a growing interest in combating the associated harmful infodemic. To this end, researchers have been curating and documenting various datasets about Covid-19. In line with existing studies, we provide an expansive collection of curated datasets to support the fight against the pandemic, especially concerning misinformation. The collection consists of 3 categories of Twitter data, information about standard practices from credible sources and a chronicle of global situation reports. We describe how to retrieve the hydrated version of the data and proffer some research problems that could be addressed using the data. 
\end{abstract}


\end{frontmatter}


\section{Introduction}
\label{sec:introduction}

Since the beginning of 2020, the dominant issue for the public is the \textit{coronavirus disease (COVID-19)}\footnote{COVID-19 and Covid-19 are used interchangeably in this work.}, which is caused by a new strain of zoonotic\footnote{i.e.~capable of being transmitted between animals and people} coronavirus (SARS-CoV-2), that was first reported by the World Health Organisation (WHO) on December 31, 2019, in Wuhan, China. The latest update from the WHO at the time of writing this paper reported 12,929,306 confirmed cases and 569,738 confirmed deaths across 216 countries, areas or territories\footnote{\url{www.who.int/emergencies/diseases/novel-coronavirus-2019}}. As cases for the Covid-19 increase, so does the associated infodemic. 
It can be argued that the need for online-based services has never been in higher demand as being witnessed in the Covid-19-induced \textit{lockdown era}. In this technology-driven society, online social networks support various forms of interactions that play a crucial role in enabling global connectivity. Because various aspects of our social lives have been affected during the pandemic with substantial reliance on online-based services, people are more vulnerable to misleading online information. While various credible bodies strive to contain the outbreak with sound and cautious approaches, various misinformation and fake news sources exist, playing a role in misleading the public. 
The proliferation of misinformation and uncensored posts on social media platforms\footnote{such as Twitter \url{www.twitter.com} and Facebook \url{www.facebook.com}} are being supported by a communication system that is focused on generating a huge amount of data (virality) at the expenses of vitality. Despite the efforts to curtail the spread of unsubstantiated claims, such as Twitter's new feature of flagging posts, it is still difficult to ascertain the veracity of the information we consume. As a result, there is huge interest at various levels to combat the pandemic.

While the scientific community is actively engaged in identifying a lasting cure for the virus causing Covid-19, i.e.~SARS-Cov-2, there is a growing interest of the computer science research community in addressing the negative impact of the corresponding infodemic. This opens up another frontier of challenge in the fight against the virus in the form of misinformation, fake news and unfounded claims. To this end, researchers have been curating and documenting various datasets about Covid-19. We observe that there is limited availability of curated ground-truth data that could be used to debunk myths and misinformation around the pandemic. Moreover, the existing tweet-based corpora, discussed in Section~\ref{sec:related-work}, need to comprise of various relevant stakeholders. 
This paper aims at addressing this gap by making available various datasets about Covid-19. The data collection\footnote{See \url{https://github.com/ijdutse/covid19-datasets} for details.} consists of 3 categories of Twitter data, information about standard practices from credible sources and a chronicle of global situation reports from WHO. The central goal of making these datasets available is to enable the study of how the spread of misinformation and rumours could be mitigated. We describe how to retrieve the hydrated version of the data and proffer some research problems that could be addressed using the data. 

\subsection{Related Work} 
\label{sec:related-work}
Within a short span of the outbreak, we have witnessed a plethora of Covid-19-related studies covering various aspects of the pandemic. We provide a cursory discussion on relevant studies and published datasets from online social networks. The interested reader may refer to \citep{latif2020leveraging}, which offers a review of related studies and a comprehensive list of relevant datasets across various domains. 
The earliest data in social networks about the virus can be traced back to January 22, 2020 by \cite{chen2020covid}. The dataset consists of tweets collected using some specific hashtags and credible accounts. The work of \cite{zarei2020first} reported a collection of image-based data about Covid-19 from Instagram. To understand the digital response in online social media, \textit{infodemics observatory} \cite{infodemic2020data} documents a large collection of public messages\footnote{see \url{https://covid19obs.fbk.eu}} related to Covid-19. The data have been classified and categorised accordingly with informative visualisations about the scale of the pandemic. The work of \cite{alqurashi2020large} provides a large collection of Covid-19 datasets in Arabic and details about Covid-19 related articles across Wikipedia projects are being documented online\footnote{\url{http://covid-data.wmflabs.org}}. Recently, Twitter made available a dedicated Application Programming Interface (API) that can be used to retrieve tweets related to Covid-19 \cite{twitter-covid-api}. 

The remaining of this paper is structured as follows: Section~\ref{sec:methodology} presents a detailed description of the data collection and processing. Section~\ref{sec:meta-analysis} offers longitudinal and exploratory analyses of the data and proffer some research problems worthy of investigation using the datasets. Section~\ref{sec:conclusion} concludes the study and discusses some future work. 

\section{The Curated Datasets}
\label{sec:methodology} 
One of the main goals of this study is to contribute diverse datasets to support research on Covid-19 and related pandemics. Thus, the onus is on identifying relevant stakeholders as the basis for the collection. 
Our datasets can be divided into two broad classes according to their sources: the \textit{tweet-based} collection, consisting of three categories of selected tweets, and the \textit{non-tweet}  collection, comprising of information about standard practices from credible sources and a chronicle of global situation reports. 
Regarding the first category, we collected Twitter data from both monitored and unmonitored accounts using relevant terms within a period of 5 weeks:  March 23 to May 13, 2020. Some of the data collection terms are captured in the following regular expressions to pull out tweets of interest. For instance, retrieving tweets matching all possible variations of referring to covid-19 using:
\begin{center}
\small
\verb^(r'[Cc]ovid(?<=)\w+|[cC]orona[vV]irus|pandemic')^
\end{center}

\subsection{Tweet-based collection}%
The tweet-based collection contains \textit{tweet objects} retrieved from Twitter via the Twitter Standard Search API based on a set of relevant keywords, as shown in Table~\ref{tab:data-sources-summary}. 
A \textit{tweet object} is a complex data object composed of numerous descriptive fields, which enables the extraction of various features for further analysis. 
Based on the method that was used for collection, the tweet-based data consist of the following data categories:

\paragraph{Account-based collection} 
Users on Twitter are broadly classified as verified or unverified. To prevent fake users masquerading celebrities or other popular individuals, Twitter performs a verification to authenticate users before appending the \textit{verified label} to the account handle. 
We monitored several accounts for a period of five weeks and retrieved tweets related to the pandemic. 
More details about the accounts that were monitored are shown in Table~\ref{tab:data-sources-summary}.

\paragraph{Random collection} 
This set consists of a generic collection of daily tweets spanning numerous topics related to the pandemic collected via Twitter's Streaming API. The rationale is to capture a wide discussion topics surrounding the prevailing pandemic. Owing to the huge amount of the retrieved tweets and noting how a hashtag represents an umbrella covering many tweets related to a topic, we retain only original posts associated with a relevant hashtag and we exclude retweets in the collection. A tweet associated with a hashtag stands a better chance of attracting much attention. This hashtag-based collection of tweets may also contain tweets in languages other than English, which have at least 100 tweets related to the pandemic.

\paragraph{Miscellaneous collection}
Noting the misinformation and myths surrounding the pandemic, we use terms associated with such myths to collect the data, as shown in Table~\ref{tab:data-sources-summary}. This category is motivated by the growing scepticism surrounding Covid-19. We manually identify users who openly dismisses Covid-19 related information put forward by credible sources such as the WHO. On that basis, each data is categorised to reflect the inclination of the user based on the content and its associated sentiment. We annotate the datasets by computing the sentiment of the users associated with the posts. The data from WHO, dubbed \textit{proWHO}, and the \textit{antiWHO}, dismissing WHO's guidelines on combating Covid-19 pandemic, are the two broad sub-categories under the \textit{miscellaneous collection} that can be used for various studies. Such studies could involve community detection and a critical analysis of users perceptions about measures taken in curtailing the pandemic.

\subsection{Non-tweet collection}
There is no gainsay that well-documented information about Covid-19 has been and is still made publicly available. 
Such endeavour is work in progress as new insights are being found. 
Consequently, we report a curated collection of datasets from credible sources concerned with combating the pandemic. 
We focus on the following sources: the World Health Organisation (WHO), UK's National Health Service (NHS-UK), the Nigeria Centre for Disease Control (NCDC-NG), and USA's Centre for Disease Control and Prevention (CDC-USA). 
The collection aims to provide informative material for a robust factual analysis and broad scope comparison. 
Through a combination of manual processing and web scrapping\footnote{using the \textit{BeautifulSoup} package, available at \url{www.crummy.com/software/BeautifulSoup}}, we obtain useful data from applicable websites. 
The rationale of using these datasets is to expand the analysis scope and enable researchers to find responses to a wide range of questions related to \textit{Covid-19}. 
While all the sources have similar content, the data from the \textit{WHO} consist of both standard practices\footnote{This ranges from debunking myths to protective measures.} and a chronicle of global situation reports.

    \begin{table}[t]
        \small
        \caption{Summary of data collection sources and corresponding description}
        \label{tab:data-sources-summary}
        \begin{tabular}{p{1.65cm} p{11cm}}
        \hline
        \textbf{Collection 1}   & \textit{Account-based collection}    \\
        \textbf{Data Source}    & Twitter   \\
        \textbf{Description}    & Tweets from specific accounts collected via Twitter's Standard Search API \\
        \textbf{Keywords}   & 
                @WHO, @DrTedros, @POTUS, @CDCgov, @ECDC\_EU, @EU\_Health,
                ECDC\_Outbreaks, @WHO\_Europe,@WhiteHouse, @HHSGov, @WHO\_Africa, 
                @NHSEngland, @NHSEnglandLDN, @officialkarimia, @metpoliceuk, @NCDCgov, 
                @WHONigeria, @fmohnigeria, @PIBFactCheck, \#HealthForAll, \#COVID 19 \\ \hline
                
        \textbf{Collection 2}   & \textit{Random collection}    \\
        \textbf{Data Source}    & Twitter   \\
        \textbf{Description}    & tweets from random accounts collected via Twitter's Streaming API \\
        \textbf{Keywords}   & 
                COVID-19, covid-19, corona virus,coronavirus, corona \& Corona, pandemic, 
                endemic, quarantine, global health, self isolate, symptoms, corona outbreak, disease, mental health, test result \& social distance, sport, 
                election politicians \\ \hline
                
        \textbf{Collection 3}   & \textit{Miscellaneous collection}    \\
        \textbf{Data Source}    & Twitter   \\
        \textbf{Description}    & A wide range collection of covid-19 data and opposition to WHO's stance \\ 
        \textbf{Keywords}   & 
                @Jordan\_Sather\_ , @kciparrish, CDC, 5G Network, 
                broadband, Wuhan China, US military, US Military Wuhan, wash hands, 
                garlic + coronavirus, mask, ventilators, Covid-19 + Bio-weapon,
                bio-weapon+covid-19, fennel cure, hydroxychloroquine, \#COVID19; \textit{we query the following accounts as the seed users:}
                @PlanB1975, @simondolan, @LonsdaleKeith, @angiebUK, @HotelLubyanka, @jcho710 \\ \hline 
        \textbf{Collection 4}   & \textit{Non-tweet-based collection}    \\
        \textbf{Data Source}    & \textit{WHO, NHS-UK, NCDC-NGN, and CDC-US}    \\ \hline
        \end{tabular}
    \end{table}
The seed users in Table~\ref{tab:data-sources-summary} were used to further identify users who share common interest or opinion with the seed users. A survey of the accounts of the seed users depicts an open dismissal of WHO's guidance on combating the pandemic.

\paragraph{Getting the hydrated tweets} 
The full version of the tweet-based collections cannot be made available due to Twitter’s policy on content redistribution\footnote{see \url{https://developer.twitter.com/en/developer-terms/agreement-and-policy}}. 
Instead of the full tweets, we provide the relevant IDs, that can be used to retrieve the data, and some meta-information about each tweet\footnote{The data presented in this paper are available at \url{https://github.com/ijdutse/covid19-datasets}}. Moreover, we include a short \textit{Jupyter notebook} with a description of how to retrieve the hydrated data and how it can be transformed onto a usable format. Alternatively, interested users may use the \textit{hydrator package} \cite{hydrator2020}.

\section{Meta-analysis}
\label{sec:meta-analysis}
This section presents some longitudinal and exploratory analyses of the data including a pointer to relevant research problems worthy of investigation using the datasets.

\paragraph{Data pre-processing} Due to the multifaceted nature of the datasets, we use the following approach to transform them so that they all conform to a unified format for redistribution. 
The tweet-based collections are messy and difficult to use directly \cite{inuwa2018detection}, hence the need for data cleaning. We carried out a rudimentary cleaning process involving tokenisation, stopwords removal and text formatting. 
To separate \textit{content words} from \textit{function words} and improve analysis, we normalise each contracted term to its expanded version. Finally, lemmatisation, the process of converting the inflected form of words back to their base forms, is performed on the normalised content words. 

    \begin{figure}[!b]
        \centering
        \includegraphics[width=\textwidth]{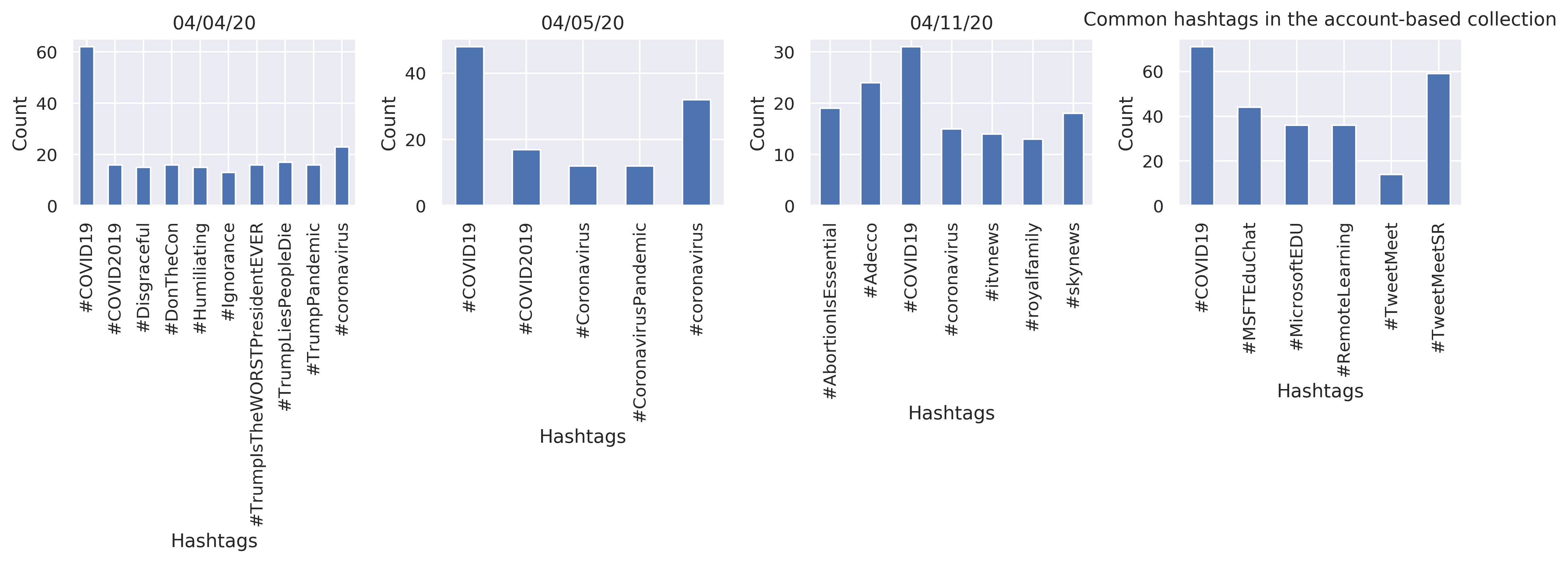}
        \vspace{-8mm}
        \caption{Most of the common hashtags in the \textit{account-based collection} relate to sensitisation and provision of genuine information about preventing the virus spread.}
        \label{fig:account-based-hashtag1}
    \end{figure}

\begin{figure}[!b]
        \centering
        \includegraphics[width=\textwidth]{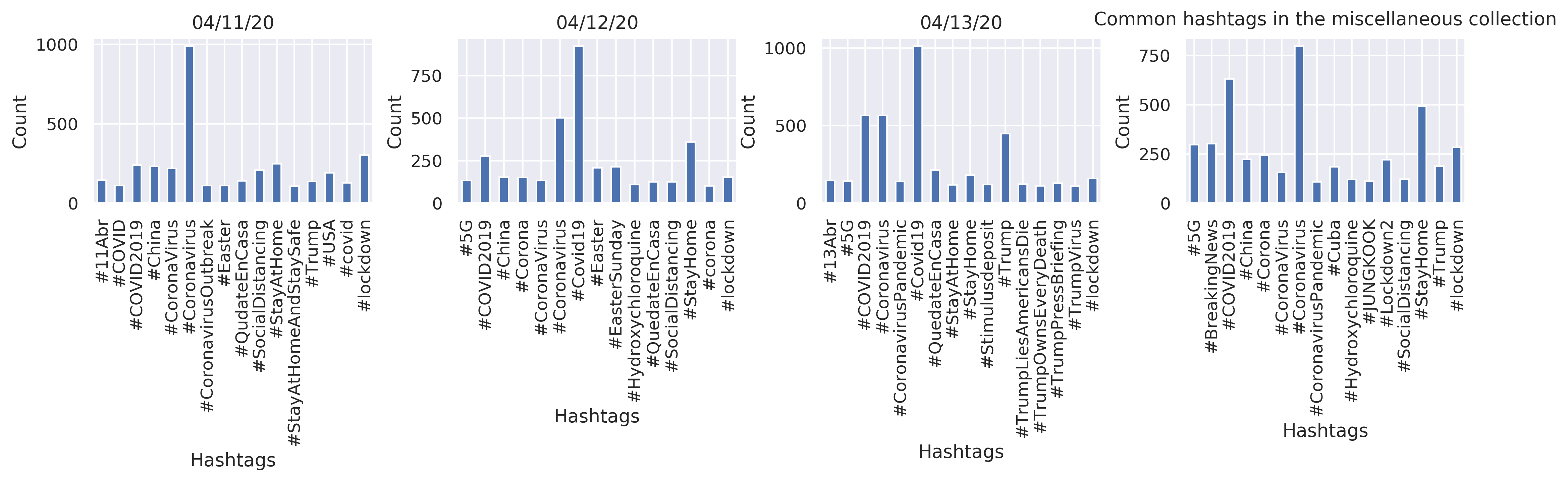}
        \vspace{-8mm}
        \caption{Common hashtags in the \textit{miscellaneous collection}}
        \label{fig:miscellaneous-hashtag1}
    \end{figure}

\paragraph{Popular accounts and hashtags} 
Figure~\ref{fig:account-based-hashtag1} shows a timeline of popular tweets over time during the collection period. The \textit{y-axis} denotes the frequency of tweets and the \textit{x-axis} denotes the popular hashtags and the corresponding dates in the title. While some of the hashtags are expected, many derogatory ones stand out in the miscellaneous collection, as shown in Figure~\ref{fig:miscellaneous-hashtag1}. With Figure~\ref{fig:account-based-hashtag1} showing the popularity of the sources or hashtags, many questions can be answered. For instance, \textit{how does the online popularity corroborate with offline relevance}? 
For the tweets posted by the monitored accounts, in the account-based collection, we determine its online presence by analysing the proportion of \textit{retweets}, \textit{favourite counts}, \textit{statuses count follower/followee} and \textit{sentiment}\footnote{We extract \textit{sentiment} using the VADER package, available at \url{https://pypi.org/project/vaderSentiment}.} associated with the texts. 

\begin{figure}[!tb]
        \centering
        \includegraphics[width=\textwidth]{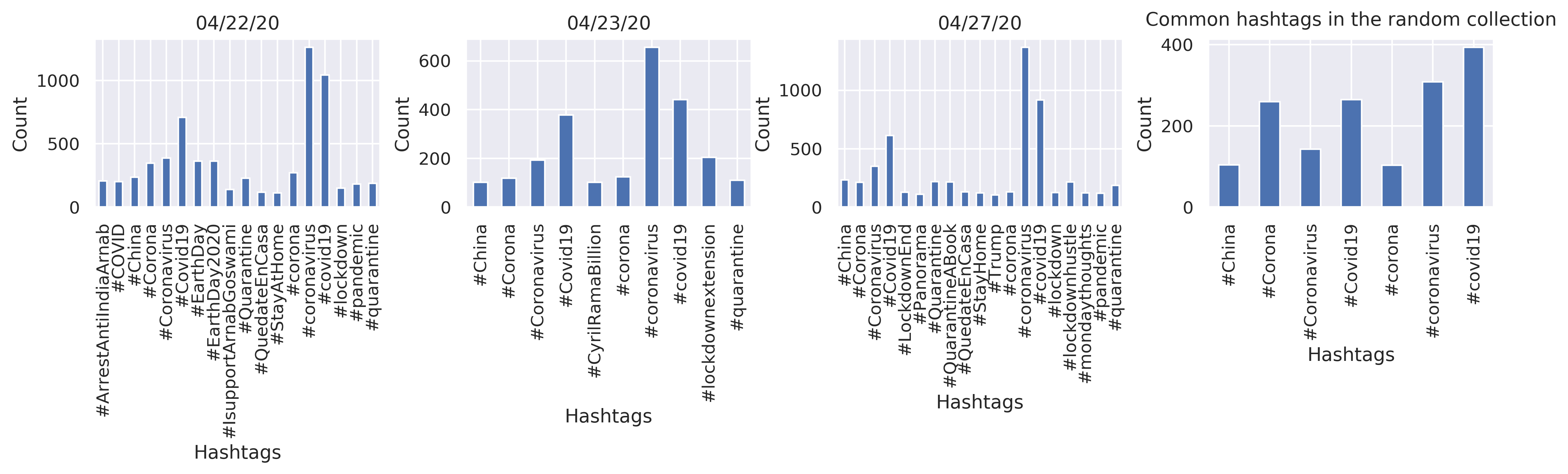}
        \vspace{-8mm}
        \caption{Common hashtags in the \textit{random collection}}
        \label{fig:random-hashtag1}
    \end{figure}

\paragraph{Ranking Hashtags} 
Because of the prevalence of many related hashtags in the collection, as demonstrated in Figure~\ref{fig:account-based-hashtag1} and Figure~\ref{fig:random-hashtag1}, we retain only the most frequent ones in the posts to enable a complete textual analysis. As a result, each tweet is merged with its corresponding hashtag text and a form of synonym resolution is performed on the hashtags. For instance \textit{\#covid-19} and \textit{\#CoronaVirus} would be similar after the resolution. 
We apply Latent Semantic Analysis (LSA) \cite{wiemer2004latent, halko2011finding}, a popular topic modelling method, to analyse further the full context of tweets in the corpus and hashtags ranking. Topic models such as the LSA and Latent Dirichlet Allocation (LDA) \citep{blei2003latent} are widely used in various tasks involving texts \citep{airoldi2008mixed,yan2013biterm,yali2014biterm}. The set of original hashtags are compared with the output of the ranked hashtag computed by the LSA and are replaced accordingly. For instance, since the LSA produces a ranked subset of the most relevant hashtags from the original set, we make a comparison and replace the set of hashtag with a single highly ranked hashtag as the best representative of the set. For those without exact representation, the first hashtag in the list is used. Figures~\ref{fig:account-based-hashtag1}, \ref{fig:random-hashtag1} and \ref{fig:miscellaneous-hashtag1} show some relevant hashtags in the tweet-based collection. 

   \begin{figure}[!tb]
        \centering
        \includegraphics[width=\textwidth]{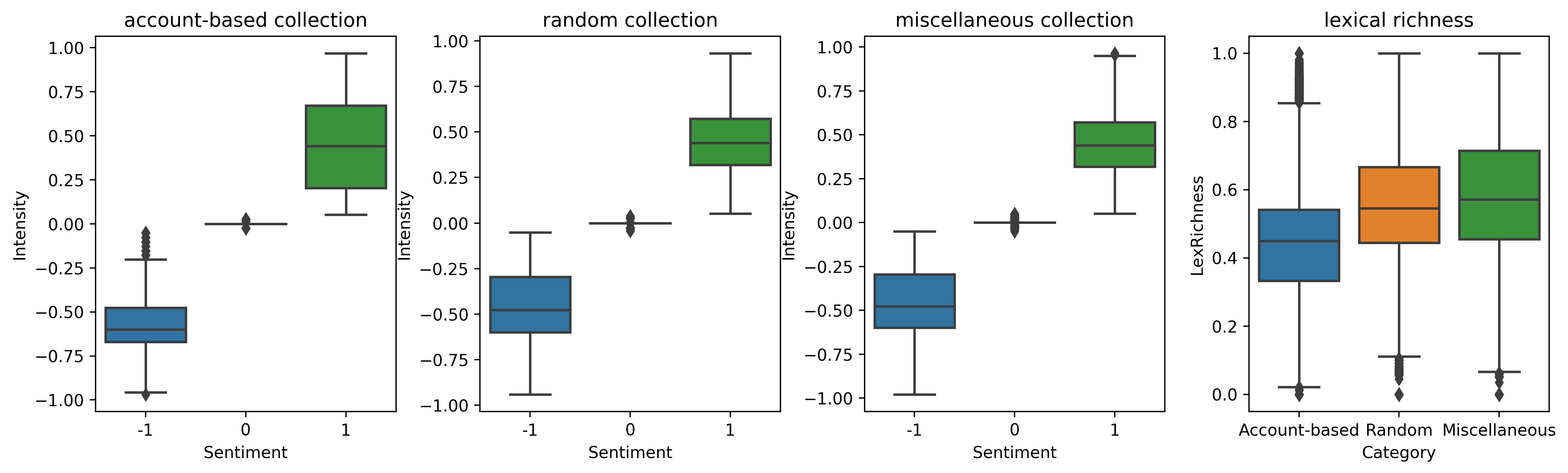}
        \vspace{-8mm}
        \caption{Sentiment and lexical richness across tweet-based collections}
        \label{fig:sentiments_and_lexrich}
    \end{figure}

\begin{figure}[!b]
        \centering
        \includegraphics[width=\textwidth]{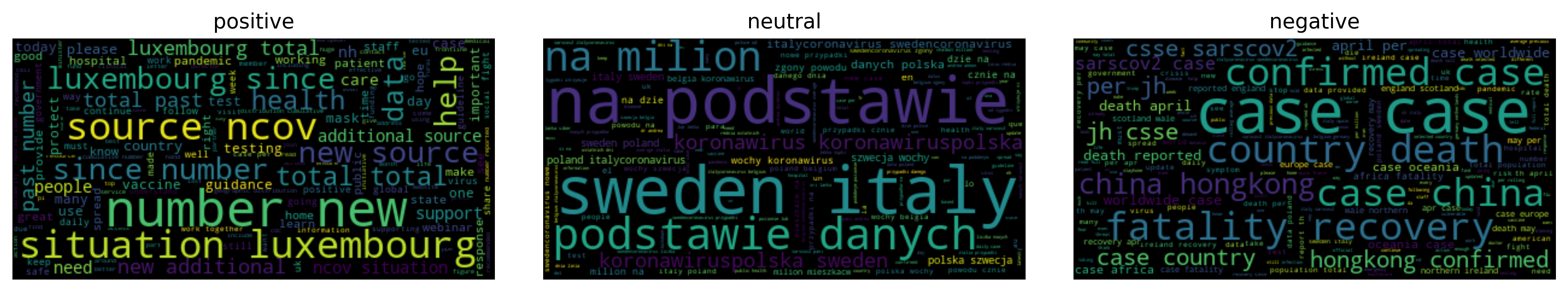}
        \vspace{-8mm}
        \caption{Sample texts in the account-based collection.}
        \label{fig:wordcloud-account-based}
    \end{figure}

\begin{figure}[!b]
        \centering
        \includegraphics[width=\textwidth]{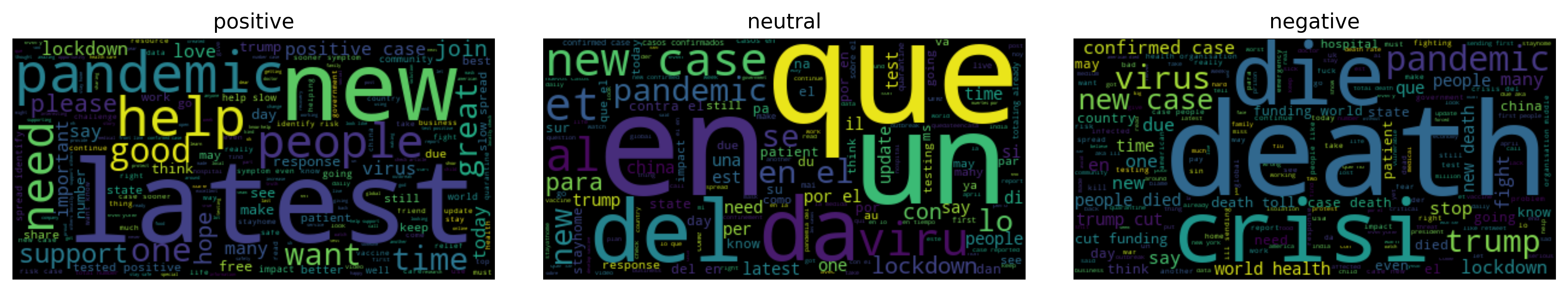}
        \vspace{-8mm}
        \caption{Sample texts in the miscellaneous set.}
        \label{fig:wordcloud-miscellaneous}
    \end{figure}

\begin{figure}[!tb]
        \centering
        \includegraphics[width=\textwidth]{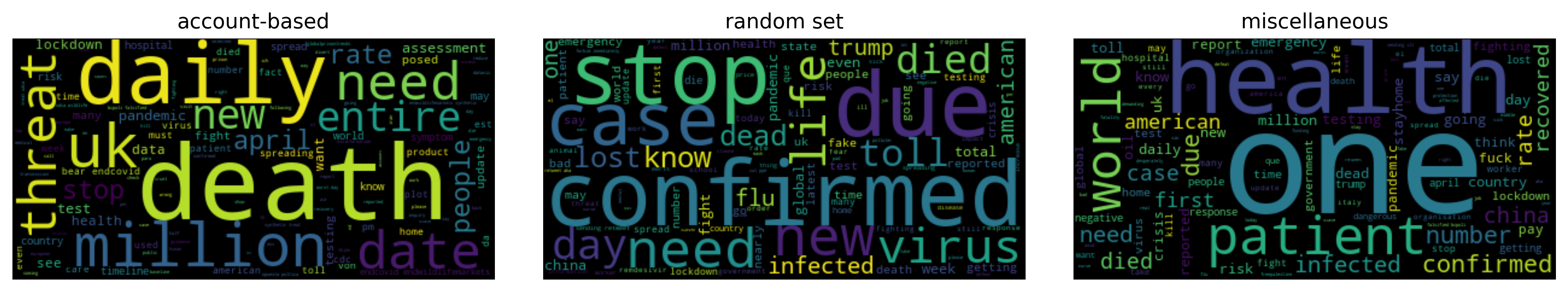}
        \vspace{-8mm}
        \caption{Sample negative terms across all collections. The visualisation is based on the learned topics using the LSA.}
        \label{fig:wordcloud-negative-lsa}
    \end{figure}

\paragraph{Relevant Themes and Sentiments} 
To identify relevant themes, we begin by aggregating a finite collection of textual content, $\mathcal{T}$, from each tweet-based collection to discover relevant themes and to analyse lexically and for sentiment. 
For any given stream of texts ${t_1, .... t_n} \in \mathcal{T}$, each $t_i \in \mathcal{T}$ consists of \textit{n-gram features}\footnote{$n$ is any positive integer, e.g.~1, 2 and 3 for \textit{unigram}s, \textit{bigrams} and \textit{trigrams}, respectively.} given by: $f_{i1}, ..., f_{im} \in t_i \in \mathcal{T}$. Accordingly, we provide a summary of the relevant themes or topics and the corresponding sentiment in Figures~\ref{fig:wordcloud-account-based} to \ref{fig:wordcloud-negative-lsa}. 
Due to the high frequency of terms related to the pandemic, e.g., covid-19 and coronavirus, we remove such variations to avoid the extinction of other relevant terms in the data. Alongside the themes, we present the \textit{sentiment} associated with the posts, reporting on different levels of polarity -- negative, neutral, positive and overall sentiment. Negative topics often relate to reports about fatalities and confirmed cases. The positive collection contains many instances about reported cases from different regions, as shown in the \textit{word clouds}\footnote{Word cloud visualisation is based on: \url{https://github.com/amueller/word_cloud}}, in Figures~\ref{fig:wordcloud-account-based} to \ref{fig:wordcloud-negative-lsa}). 

\paragraph{Lexical Richness} 
With respect to the tweet-based collections, we analyse \textit{lexical richness} to understand the diversity of lexicons used across them. Our intuition is that content irrelevant to Covid-19 will have be richer lexically, because the narration and conspiracies surrounding the pandemic keep changing, hence employing diverse wording in the discourse. On the other hand, content from recognised sources, such as the WHO, would remain relatively consistent and slow to change due to the  formal process and other health-related protocols or standard practices that apply before posting. Moreover, lexical richness can be viewed as a measure of consistency in the narration about the pandemic. A longitudinal analysis of the richness of content may provide a glimpse of what is happening. 
Figure~\ref{fig:sentiments_and_lexrich} shows both sentiment intensity, in the first three subplots, and lexical richness, in the last sub-plot, across the tweet-based collections. Indeed, the observed result confirmed our intuition, showing low richness in the account-based collection, which contains tweets that mostly come from credible sources (see Table~\ref{tab:data-sources-summary}). Figures~\ref{fig:wordcloud-account-based}, ~\ref{fig:wordcloud-miscellaneous} and ~\ref{fig:wordcloud-negative-lsa} show some examples from the \textit{topic analysis} in form of the most relevant terms in the collections. 

\subsection{Utility of the datasets}
In this section, we identify relevant areas and research problems that could be studied using the datasets. 
The increasing number of uncensored posts, partly due their short size and the speed of communication via posts on virtually anything, facilitates the proliferation of content that affects the credibility of information on social media. 
Despite the measures taken by social media platforms to curtail irrelevant content, many sources of misleading information and rumours still exist. 
Concerning the Covid-19 pandemic, there exist various misinformation and conspiracy sources capable of misleading the public. We identify the following relevant areas and high-level research problems that could be investigated using the datasets presented in this work. 
    \begin{itemize}
        \item[-] \textbf{content classification:} to distinguish genuine content in an ecosystem cluttered with spurious content
        \item[-] \textbf{content validation:} the focus is to ascertain the veracity of a tweet or document using a large collection of ground-truth data from credible sources.
        \item[-] \textbf{tone of misinformation:} of interest is to analyse the tone of misinformation in comparison to the genuine counterparts -- emotional, negative, neutral, positive.
        \item[-] \textbf{misinformation diffusion:} to understand the propagation of news, virus infection, rumour, fake news and unfounded claims. It would be worthwhile to study the propagation of information related to the pandemic and study how to optimise methods to ensure that relevant content dominates and irrelevant content is suppressed. 
        \item[-] \textbf{thematic analysis:} the use of a streaming API returns a huge collection of somewhat irrelevant data. 
        It would be useful to streamline the work to focus on a fixed number of tweets from each account and then perform textual analysis to understand the topics being discussed.
        It is worth noting that a single tweet is usually limited in conveying sufficient information about a topic, making it difficult to fully understand the discussion context. 
        Future work could be extended to understand the specific topics being discussed and compare the degrees of similarity among documents. 
        \item[-] \textbf{evaluation of lockdown policy:} It would be interesting to assess or gauge users' acceptance of lockdown measures. This would enable the quantification of various lockdown policy measures. For instance, each lockdown measure ca be associated to the sentiment of users. Furthermore, it would make sense to track sentiment over time and see how attitude changes.
        \item[-] \textbf{community detection:} the miscellaneous collection consists of two broad categories of users annotated as \textit{proWHO} and \textit{antiWHO} based on their inclinations towards the pandemic. Because this is a high-level categorisation, we postulate that many overlapping communities may exist. For instance, the \textit{proWHO} group may consists of users who dismisses the content posted by the group. This is also true in the \textit{antiWHO} group. The dataset can be used for various studies such as community detection and an indepth analysis of users' perceptions about the pandemic.
    \end{itemize}


\section{Conclusion}
\label{sec:conclusion}
The prevailing technological advancements culminated in the computerisation and automation of various tasks, which leads to a complex ecosystem of information exchange mostly via social media platforms. The architecture of social media networks simplifies the spread of information to a wide audience making it a useful facility for instant information update and socialisation; this also results in an increasing number of uncensored posts on virtually anything. The prevailing Covid-19 pandemic in the age of social media has reveal to us that as the virus propagate, so is false or misleading information about it leading to infodemic. A case in point is the various myths associated with the \textit{Covid-19}, which often left the public bewildered concerning what preventive measures to take and which information to believe. Misinformation about Covid-19 can be considered along the dimensions of the origin of the virus, the symptoms, preventive measures and cure. The latter is probably the most crucial since an ill-formed medical advice will be catastrophic. 

Because misleading information can have catastrophic consequences and hampers the fight about applying containment measures, it is pertinent to combat the  pandemic from all possible fronts. Consequently, our work contributed a curated collection of relevant datasets to support multi-faceted research works interested in the Covid-19 infodemic and related tasks. We provide a basic but useful analyses of the datasets and proffer direction/relevant areas to study using them. The datasets from various credible sources can be used as benchmark to assess the veracity of information related to the pandemic. 
The data will further enrich existing databases for debunking misinformation and fact-checking avenues, such as the International Fact-Checking Network\footnote{IFCN: \url{www.poynter.org/ifcn}} and AFP\footnote{AFP: \url{https://factcheck.afp.com}}. The datasets can help in understanding how to optimise or ensure that relevant content dominates and irrelevant content is suppressed, especially during critical times of the pandemic. 

\section*{Acknowledgements}

The second author has participated in this research work as part of the TYPHON Project, which has received funding from the European Union’s Horizon 2020 Research and Innovation Programme under grant agreement No.~780251.


\bibliographystyle{unsrt}
\bibliography{covid-19-datasets}
\end{document}